\documentclass[twocolumn]{aastex62}

\graphicspath{{./}{figures/}}

\begin{document}

\title{A light redback companion of PSR J1622$-$0315 and irradiation power in spider systems}

\author{Y. X. Jane Yap}
\affil{Institute of Astronomy, National Tsing Hua University, Hsinchu 30013, Taiwan \textbf{R.O.C}}

\author{Albert K. H. Kong}
\affil{Institute of Astronomy, National Tsing Hua University, Hsinchu 30013, Taiwan \textbf{R.O.C}}

\author{Kwan-Lok Li}
\affil{Department of Physics, National Cheng Kung University, Tainan 701, Taiwan \textbf{R.O.C}}




\begin{abstract}

We report optical observations of 
  the millisecond pulsar binary system PSR J1622$-$0315 
  with the Lulin 1m telescope in Taiwan and the Lijiang 2.4m telescope in China 
  between 2019 and 2021. 
The companion of the pulsar, which is of V$\sim$19~mag, 
  showed ellipsoidal-distorted orbital variations in its light curves. 
The best-fit model to the light curves,  
  with the binary code {\tt PHOEBE}, gives 
  a companion mass of 0.122$\pm 0.006\;\! {\rm M}_\odot$.  
This places PSR J1622$-$0315 in the spider-system subclass. 
We compared the properties of PSR J1622$-$0315 
  with other spider pulsar binaries  
  for the scalings between the spin-down luminosity derived for the pulsar,  
  irradiation luminosity of the companion, and 
  X-ray luminosity of the binary. 
We find that pulsar irradiation in PSR J1622$-$0315 is insignificant 
  and the irradiation luminosity of the transitional millisecond pulsars 
  PSR J1023$+$0038 and PSR J1227$-$4853 
  are the highest among the redback systems.

\end{abstract}





\section{Introduction} 
\label{sec:intro}


Spider pulsar systems are compact binaries containing 
  a millisecond pulsar (MSP), with a low-mass companion star,  
  orbiting around each other in a period of $P_{\rm b}\lesssim 24~{\rm hr}$.  
They are usually classified as black widows (BWs) or redbacks (RBs).  
The companion stars generally have masses $\lesssim 0.1\;\! {\rm M}_\odot$ and $\sim 0.1-0.4\;\! {\rm M}_\odot$  for BW and RB, respectively
\citep[see e.g.][]{Chen2013, Roberts2013}.
A rapid-spinning MSP is believed to be a phenomenon caused by 
  the accretion of material from the companion star\footnote{known as the recycling scenario \citep{1982Natur.300..728A}}.    
This scenario is supported by observations 
  which showed state transition(s) 
  between accretion-powered and rotation-powered (pulsar) state 
  in three transitional MSPs (tMSPs): PSR J1023$+$0038 \citep{Archibald2009}, PSR J1227$-$4853 \citep{Bassa2014} and PSR J1824$-$2452I \citep{Papitto2013}. 

The light curve of the companion star in spider systems contains information about the irradiation of the system and the companion's stellar properties.
The effect of strong irradiation is observed in a few RBs and BWs
  \citep[e.g.,][]{Breton2013, Draghis2019}.
Pulsar irradiation also causes evaporation of the companion star and 
  results in the mass loss of the star \citep{EPJVanH1988}.
The evolution history of a companion star in spider systems evolving from a low-mass X-ray 
  binary (LMXB) system under ablation was discussed in \cite{Chen2013}.
The interaction between the magnetodipole radiation and a disk was proposed to explain the neutron star rotation properties in these systems, and the model was applied in transitional systems during the rotation-powered state. \citep{Burderi2001,Papitto2015}. 

In this work, we report new optical observations of PSR J1622$-$0315. 
We use Markov Chain Monte Carlo (MCMC) sampling to explore the parameter space in the models we use to fit the light curves, and to estimate the uncertainties
of the masses of the components and the orbital properties of the system.
We constructed a table of archival values to compare the spider systems in terms of their irradiation luminosities ($L_{\rm irr}$) inferred from the light curves, X-ray luminosities ($L_{\rm x}$) of the systems, and spin-down luminosities ($\dot{E}$) of the pulsars.

We introduce our target and summarize our observations in Section~\ref{1622}.
In Section \ref{sec:phoebe_model}, we describe our light curve analysis 
  using the eclipsing binary modeling code {\tt PHOEBE} \citep{Prsa2016}.
In Section \ref{sec:MWL}, we discuss 
  the empirical relationship between $L_{\rm irr}$, $L_{\rm x}$ and $\dot{E}$ in BWs, RBs and tMSPs.
  We summarize and discuss the implications of this work in Section 5.



\renewcommand{\arraystretch}{1.3}
\begin{table}
    \caption{Observation details of MSP J1622$-$0315}
    \centering
    \begin{tabular}{p{0.18\columnwidth} p{0.27\columnwidth} p{0.1\columnwidth} p{0.15\columnwidth} p{0.12\columnwidth} }
        \hline\hline 
        Telescope & Date & filter (SDSS) & exposure time (s) & duration (mins) \\
        \hline
        Lulin-1m & 2019 February 2 & $r',~g'$ & 180, 300  & 120\\
        & 2019 February 3 &$r',~g'$ & 180, 300  & 120\\
        & 2019 April 6 &$r',~g'$ & 180, 300  & 180\\
        & 2021 July 4$^{\dagger}$ &$r',~g'$ & 180, 300  & 36\\
        Lijiang-2.4m  & 2019 March 10 &$r',~g'$ & 180, 300  & 80\\
                \hline 
    \end{tabular}
    \tiny\scalebox{1}{}{ 
    \begin{flushleft}
    $^{\dagger}$ Opportunity allowed for the collection of additional quality data on July 4, 2021 with the LOT for slightly more than half an hour.
    \end{flushleft}}
    \label{tab:obs}
\end{table}



\section{Target and observations}
\label{1622}

\subsection{PSR J1622$-$0315}
PSR J1622$-$0315 is a binary MSP discovered by \cite{Sanpa} using the Green 
  Bank Telescope (GBT) and the Nan\c{c}ay telescope. 
It has a spin period of 3.845~ms, and
it is located at coordinates RA=16\textsuperscript{h}22\textsuperscript{m}59\textsuperscript{s}.6 and DEC=-03\degr15\arcmin37.3\arcsec (J2000).
The system has an orbital period of 3.9~hr and the dispersion measure is 
  21.4pc~cm$^{-3}$ ($\sim$20\% uncertainties; \citealt{Sanpa}).
Radio flux variations, possibly caused by scattering with the ISM or ejected 
  particles from the companion were also reported~\citep{Sanpa}. 
An optical counterpart was identified with the MDM Observatory located in 
  Arizona, United States, and its light curve exhibits an ellipsoidal variation. From the pulsar timing, the mass function was obtained and a companion mass of $>$0.1$\;\! {\rm M}_\odot$ was derived, assuming a neutron star (NS) mass of 1.35$\;\! {\rm M}_\odot$ and an edge-on orbit. 
No strong emission lines were observed in the optical spectrum of the 
  companion of PSR J1622$-$0315 \citep{Strader19}.
Measurements of the companion radial velocity further allowed 
  the neutron star mass to be constrained to $>$1.45$\;\! {\rm M}_\odot$ for an edge-on orbit or, alternatively, constrained the inclination to $>$64\degr~if the pulsar is less massive than 2.0$\;\! {\rm M}_\odot$.
\cite{Gentile2018PhD} studied the X-ray light curve and spectrum of PSR 
  J1622$-$0315 with \textit{XMM-Newton}. Although the X-ray count rate was the lowest at the superior conjunction of the pulsar, the orbital modulation was not obvious. 
The X-ray spectrum was fitted with a power law of photon index $\Gamma$=2.0 
  and there was no strong indication of a thermal component.
Additionally, folded \textit{Fermi-\rm {LAT}} observation indicated plausible 
$\gamma$-ray pulsation \citep{Sanpa}.


\subsection{Observations and timing analysis}
\label{sec:obs}

We performed a series of targeted observations using the Lulin 1-m telescope 
  (LOT) in Taiwan and the Lijiang 2.4m telescope in mainland China during the first quarter of 2019 (complemented with a shorter epoch obtained in July 2021). We used the SDSS r' and g' filters, and alternated them between each exposure.
The details of the observations are listed in Table \ref{tab:obs}.
The combination of our observations allowed us to cover all orbital 
  phases, with data spanning over a few months.
We only analyzed the observations which occurred when the weather conditions allowed a seeing better than FWHM$\lesssim$~3\arcsec.
The reported light curves include 68$\times$180~s exposures of ${r'}$ images 
  and 66$\times$300~s exposures of ${g'}$ images.
Raw images were calibrated with bias and flat frames using the {\tt IRAF V2.15}
  standard pipelines. 
We performed photometry for all image frames using the {\tt IRAF} package {\tt PHOT}.
We selected eight spherical, unsaturated point sources that are relatively isolated, as comparison stars, for all image frames.
Source counts were extracted from circular apertures with 4\arcsec~radii for the target and comparison stars in the field. We extracted background counts from annular regions centered at the positions of the stars, with inner radii of 8\arcsec~and width of 2\arcsec. The photometry was performed using constant weighting, and the background rate was calculated using the centroid sky fitting algorithm.
We then computed the differential photometry of PSR J1622-0315 by subtracting the average magnitudes of the comparison stars from the instrumental magnitudes of the target. The propagated errors of the measurements are shown in the error bars.

To convert the differential magnitude into apparent magnitude, we selected another stable bright star\footnote{GAIA EDR3 ID 4358429221667653632} as a reference star and calculated its differential magnitudes with the same differential photometry process described above. The differential magnitudes of the reference star varied less than 0.1 mag among the frames, which is sufficient for magnitude calibration. 
  The apparent magnitude of the reference star is obtained from the Sloan Digital Sky Survey (SDSS) Data Release 14 catalog.
  The difference between the differential magnitude and the apparent magnitude of the reference star is used to calibrate the light curve of PSR J1622-0315.

\begin{figure}
    \centering
    \includegraphics[scale=0.5]{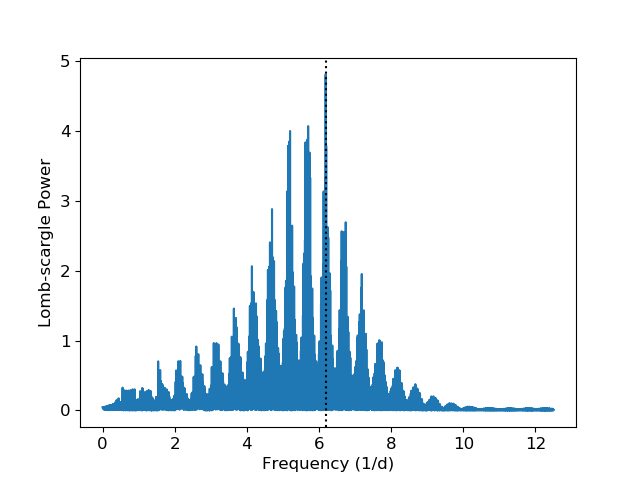}
    \caption{
    Lomb-Scargle periodogram of the combined light curves taken in 2019 for PSR J1622$-$0315.
    The peak with the strongest amplitude is located at a frequency that corresponds to a period of 0.162(2) days, indicating the orbital period of the system.
    Our result is consistent with the previously reported orbital period obtained using radio timing, which is indicated by a dotted line.
    }
    \label{fig:lomb}
\end{figure}

\begin{figure}
    \includegraphics[scale=0.4]{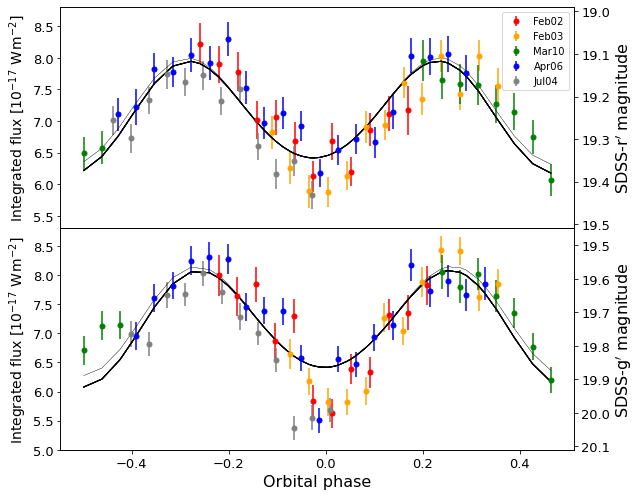}
    \caption{The folded light curve of PSR J1622-0315 companion in ${r'}$(top) and ${g'}$(bottom) band. Phase 0 corresponds to the inferior conjunction of the companion. 
    The black solid line is the best-fit model from PHOEBE. The most recent dataset
(2021 July 4) is not included in the fit. 
    The gray solid line demonstrates the effect of irradiation on the companion star surface due to heating from the pulsar; the effect is most obvious at the companion superior conjunction where most of the extra irradiation is seen.}
    \label{fig:p1}
\end{figure}

The date is converted to Barycentric Julian Day (BJD) time before performing 
  timing analysis. We performed a Lomb-scargle periodogram analysis on the selected light curves using {\tt astropy.timeseries}. We found a period of 3.88(5)~hr in the light curves (c.f. Figure \ref{fig:lomb}).
This period is consistent with the more precise pulsar timing 
  solution previously reported (0.1617006798~d; \citealt{Sanpa}), therefore we adopted the latter in the following modeling analysis. 
We use $T_0$ (ascending node of the companion) published in \cite{Strader19} to fold the data.
The light curve obtained is shown in Figure \ref{fig:p1}.



\begin{figure}
    \centering
    \includegraphics[scale=0.45]{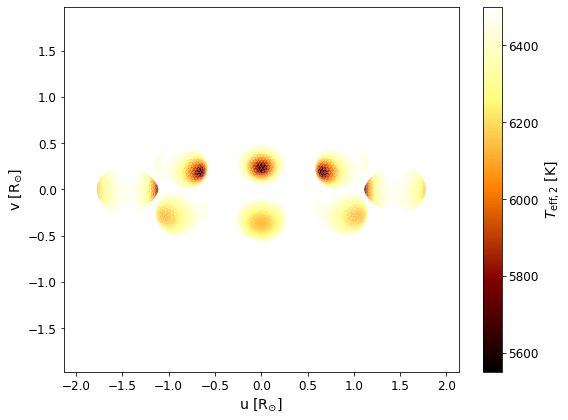}
    \caption{Projected view of the companion star across different orbital phases, including its effective temperature distribution. This model corresponds to the parameter values reported in Table 3.}
    \label{fig:phoebe}
\end{figure}



\section{Light curve modeling}
\label{sec:phoebe_model}
We conducted an inverse modeling analysis following the 
  method described in \cite{Conroy2020} using {\tt PHOEBE 2.3}.
  To begin, the apparent magnitudes are converted to the integrated fluxes using
  the asinh softening parameters and zero points provided in the SDSS documentation.\footnote{\url{ http://classic.sdss.org/dr6/algorithms/fluxcal.html}}
The ${r'}$ and ${g'}$ light curves are loaded into {\tt PHOEBE} as separate 
  light curve datasets with the passband set to their respective colors. 
We used data taken between February and April 2019, which covered the entire 
  orbit for our analysis. The most recent dataset (2021 July 4) is not included in the modeling because it was taken substantially later compared to other datasets, to avoid possible changes in the system during this period. 
The position of the target with respect to the location of the telescope 
  prevented us from achieving full orbital coverage in a single observing session of one night. While there were fluctuations between sessions, the overall ellipsoidal modulation is clearly seen in the folded light curves.
We fixed the orbital period (\textit{P}$_{\rm b}$) and the superior conjunction of the pulsar ($T_{0,\rm {sup\_conj}}$)
using the adopted values mentioned in Sections~\ref{sec:obs}, with an additional phase shift of 0.25 to account for the different definition of phase 0 used in {\tt PHOEBE}. We also adopted the projected semi-major axis of 0.219258 lt-s from \cite{Sanpa}.

We modeled PSR J1622$-$0315 to resemble a semi-detached binary that consists 
  of one primary (subscript `1'; pulsar) and one secondary (subscript `2'; companion) component, where the companion star has a radius approaching the maximum allowed value before overflowing the Roche-lobe. 
We assumed no eclipse in the system (``\textit{eclipse method}"=``\textit{only 
  horizon}"), since the light curve variation mainly comes from ellipsoidal distortion.
The passband luminosity mode is set to ``\textit{absolute}'' to avoid any flux 
  rescaling. For the passband limb-darkening correction, we use a logarithmic law for the secondary star, and the coefficient is derived from the atmosphere model \citep{ck2004}
``\textit{ck2004}'' implemented in {\tt PHOEBE}. 
Since the optical emission from a pulsar is negligible, we can turn off the primary star contribution
by setting ``\textit{distortion method}" to ``\textit{none}".
We also assumed a convective envelope for the companion star where the 
  bolometric gravity darkening coefficient $\beta$ is 0.08. This value is also used in the studies of similar systems~\citep[e.g.][]{Bellm2016}.
The modeled system and the temperature distribution of the companion is shown 
  in Figure \ref{fig:phoebe}.
  The inner tip of the companion star appears coldest compared to the rest of the surface; this is because the effective gravity near the tip is among the lowest under the influence of the tidal effect.
  The inner tip would appear brighter if significant pulsar irradiation were present.

Figure \ref{fig:p1} also illustrates a light curve (in gray) with irradiation on the surface of the companion star caused by the pulsar. 
 This is done by setting the irradiation method to ``\textit{wilson}" \citep[]{Wilson1990}. 
For the primary star inputs, we changed the distortion method back to ``\textit{roche}", and applied 
the effective temperature and the radius of a neutron star to allow the irradiation effect to take place.
The flux of the primary star heats the surface of the secondary star, and results in an increase of temperature, especially near the inner tip of the secondary star.
As shown in the light curve, the irradiation in this system is not significant (the difference in the computed loglikehoood is smaller than 0.01\%); if added to the model, the effect is mostly observed at the superior conjunction of the companion star. A more severe irradiation will not reproduce the data well. Due to the uncertainty in the irradiation physics for spider MSPs, this effect
is not included in the following MCMC analysis.


\subsection{MCMC sampling}
\label{mcmc}
We performed a MCMC study
  for six system parameters: mass ratio (\textit{q\textsubscript{\rm binary}}; $M$\textsubscript{comp}/$M$\textsubscript{NS}), companion's effective temperature (\textit{T\textsubscript{\rm eff,2}}), system inclination (\textit{i\textsubscript{\rm binary}}), companion radius (\textit{R\textsubscript{\rm equiv,2}}), extinction parameter (\textit{A\textsubscript{\rm V}}) and distance (\textit{d}). 
We used the \texttt{EMCEE} \citep{emcee2019} solver implemented in {\tt PHOEBE}.
We note that although the filling factor is more commonly used in the spider 
  community, it is not a default parameter for detached binary models in {\tt PHOEBE}. However, a volume-averaged filling factor can be derived using the critical Roche radius provided by the model. 
 
We assigned uniform priors for \textit{q\textsubscript{\rm{binary}}, T\textsubscript{\rm eff,2}, i\textsubscript{\rm binary}, A\textsubscript{\rm V}} and \textit{d}, and a Gaussian prior for \textit{R\textsubscript{\rm equiv,2}}. 
The ranges for each of the uniform priors are shown in Table \ref{tab:prior}.
We chose a uniform prior for \textit{q\textsubscript{\rm{binary}}} 
with limits ranging from 0.04 to 0.12 for pulsar mass ranges from 1.4 \({\rm M}_\odot\) to 2.0 \({\rm M}_\odot\).
Based on the binary mass function value of 1.27 \({\rm M}_\odot\) \citep{Strader19}, we set the lower limit for the system inclination at 50$\degr$ to avoid fitting a large neutron star mass ($>$3 \({\rm M}_\odot\)).
For \textit{T\textsubscript{\rm eff,2}}, the limits are chosen based on the wide range of temperatures observed in RB companions. 
  The $E$(B$-$V) color excess is 0.25 {mag} at the given coordinate and 
  distance, by examining the Bayestar3D dust map through the python package \texttt{dustmaps} \citep{2019ApJ...887...93G}. 
The color excess is then converted to \textit{A\textsubscript{\rm V}} using $R_{\rm V}= 3.1$. 
We set a uniform prior between 0.7 and 0.9~{mag}.
Due to the degeneracy between \textit{A\textsubscript{\rm V}}, \textit{T\textsubscript{\rm eff,2}} and \textit{d}, it is necessary to incorporate more than one color information in order to achieve a better fit.
The distance obtained from \cite{Gaia2021} is 5843 pc with uncertainties range 
  from 1664 pc to 7766 pc. 
For \textit{R\textsubscript{\rm equiv,2}}, we use a Gaussian prior with a mean value of
  0.27$\;\! {\rm R}_\odot$, which corresponds to roughly 90$\%$ of the critical Roche radius.
 
  An initial sample of fitting parameters is drawn from Gaussian distributions centered around the mean of the priors.
The initial sample distributions and prior distributions are chosen such that the desired parameter space can 
  be fully explored within a reasonable computational time. 
We exposed the failed walkers (due to physical or backend limitations) to make sure that the parameter space is successfully covered. 
As demonstrated in \cite{Conroy2020},
these limitations are used to constrain the parameter search space if no priors are provided explicitly.


\renewcommand{\arraystretch}{1.4}
\begin{table}
    \centering
    \begin{tabular}{ccc}
    \hline\hline
       Lower limit  & Parameter & Upper limit \\
    \hline
       0.04  & \textit{q\textsubscript{\rm{binary}}} & 0.12\\
         3500 & \textit{T\textsubscript{\textup{eff,2}}} ({K}) & 7000\\
         50 & \textit{i\textsubscript{\textup{binary}}} ($^{\circ}$) & 90\\
         0.7 & \textit{A\textsubscript{\rm V}} ({mag}) & 0.9 \\
         1.664 & \textit{d} ({kpc}) & 7.766\\
    \hline
    \end{tabular}
    \caption{Uniform priors are used for the above fitting parameters. The prior limits are discussed in the text. }
    \label{tab:prior}
\end{table}



\begin{figure*}
    \centering
    \includegraphics[scale=0.5]{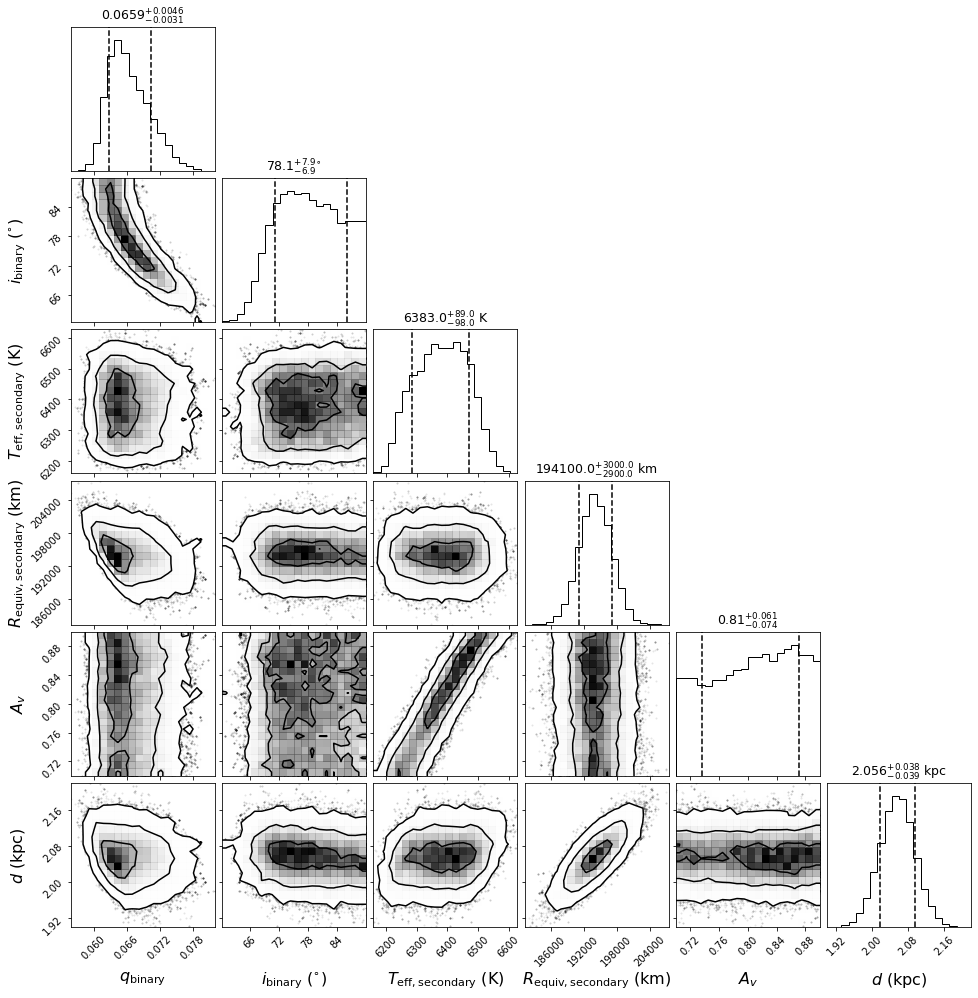}
    \caption{Corner plot of the posterior distributions for the fitted parameters. 
    The median values are displayed with the respective 1$\sigma$ uncertainties, indicated by dotted lines. 
    Degeneracy is observed in the following parameters: i) $q_{\rm binary}$ and $i_{\rm binary}$ ii) $T_{\rm eff, secondary}$ and $A_{\rm V}$, iii) $R_{\rm secondary}$ and $d$. Given the data quality, we were unable to constrain the extinction parameter, $A_{\rm V}$.} 
    \label{fig:mc}
    
\end{figure*}

\begin{figure}
    \centering
    \includegraphics[scale=0.5]{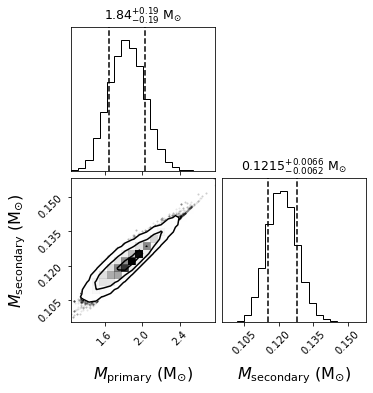}
    \caption{Propagated distribution of parameters $M_{\rm primary}$ (NS) and $M_{\rm secondary}$ (companion). The median values are displayed with the respective 1$\sigma$ uncertainties.}
    \label{fig:mass}
\end{figure}



\subsection{MCMC results}
We ran 30 walkers with 10600 iterations, where around 3000 steps were dropped as burn-in. 
Figure \ref{fig:mc} shows the corner plot of the posterior distributions.
We obtained a mass ratio of $\sim$0.066, which is similar to that reported in 
  \cite{Strader19} using a radial velocity measurement. 
The existing degeneracy between the mass ratio and the inclination parameter is observed.
The sampled density for the companion's effective temperature peaks around 6400~{K}. 
We note that the fitted effective temperature is higher than most known RB companions 
  ($\sim$3300$-$5500 {K}; see references in Table \ref{tab:profile}). 
  The extinction parameter, $A_{\rm V}$, is not well constrained due to the large uncertainties in our data.
We obtained a system distance of $\sim$2.1~{kpc} and a companion star radius  of $\sim$0.28~${\rm R}_{\odot}$.
The determination of the distance of the system is affected by the companion star radius parameter. If the radius increases, a larger distance is required to fit the light curve.
The maximum Roche radius, as a result of the orbital parameters and the mass ratio, is 0.283 ${\rm R}_{\odot}$. Hence, the volume-averaged filling factor is close to 1.
A semi-detached system with a close-to-unity filling factor is common among spider 
  systems \citep{Vito2020}.
The binary masses are derived from the fitted mass ratio and inclination
  (c.f. Figure \ref{fig:mass}). 
The mass ratio implies a 1.84-$\textup{M}_\odot$ pulsar and a 0.122-$\textup{M}_\odot$ 
  companion with a high inclination of $\sim$80$\degr$. 
All best-fit parameters are shown in Table \ref{tab:3}.
This result suggests that the PSR J1622$-$0315 companion is the lightest RB companion 
  known so far.



\renewcommand{\arraystretch}{1.4}
\begin{table}
    \centering
    \begin{tabular}{p{0.55\linewidth} p{0.25\columnwidth}}
    \hline\hline 
    Fit parameter(s) & Result(s)  \\
    \hline
         \textit{q\textsubscript{\rm{binary}}} [$M$\textsubscript{comp}/$M$\textsubscript{NS}]& $0.066_{-0.003}^{+0.005}$\\
         \textit{T\textsubscript{\textup{eff,2}}} ({K}) & $6383_{-98}^{+89}$ \\
         \textit{i\textsubscript{\textup{binary}}} ($^{\circ}$) & $78.1_{-6.9}^{+7.9}$  \\
         \textit{R\textsubscript{\textup{equiv,2}}} ({\({\rm R}_\odot\)})
         &  $0.279_{-0.004}^{+0.004}$ \\
         \textit{A\textsubscript{\rm V}} (mag) & $0.81_{-0.07}^{+0.06}$\\
         \textit{d}~(kpc) & $2.06_{-0.04}^{+0.04}$\\
         \hline
         Derived parameter(s)&\\
         \hline
         \textit{M\textsubscript{\rm primary}} (\({\rm M}_\odot\))& $1.84_{-0.19}^{+0.19}$ \\
         \textit{M\textsubscript{\rm secondary}} (\({\rm M}_\odot\))& $0.122_{-0.006}^{+0.007}$\\
         \hline
    \end{tabular}
    \caption{Best-fit results with 1$\sigma$ uncertainties.}
    \label{tab:3}
\end{table}
\renewcommand{\arraystretch}{1}
\begin{table*}
\small
    \centering
    \begin{tabular}{p{0.03\textwidth} p{0.15\textwidth} p{0.1\textwidth} p{0.055\textwidth} p{0.07\textwidth} p{0.11\textwidth} p{0.1\textwidth} p{0.1\textwidth} p{0.07\textwidth} p{0.1\textwidth}}
    \hline\hline
    Item & Designation & BW/RB (\textsuperscript{a}) & \textit{P}~ & $\dot{P}$& Companion  &$L_{\rm irr}~$     &$ \log_{10}\;\!(L_{\rm x}/$ & $\log_{10}\;\!(\dot{E}/$ & References \\
    &&&(ms)&($\times 10^{-21}$ s\;\!s$^{-1}$)&mass (${\rm M}_\odot$)&($\times 10^{34}$ \;\!erg\;s$^{-1}$)&erg\;s$^{-1}$)&erg\;s$^{-1}$)&\\
        
\hline
1 &PSR J0023$+$0923	&BW	(1)& 3.05& 9.61&$0.018^{+0.002}_{-0.001}$	&$0.296^{+0.016}_{-0.013}$ &$29.35^{+0.47}_{-0.53}$	&34.13	& 1,2 \\
2 &PSR J0251$+$2606	&BW	(1)& 2.54& 7.57&$0.032^{+0.006}_{-0.003}$	&$0.252^{+0.012}_{-0.011}$ &-	&34.26 &1,4  \\
3 &PSR J0636$+$5128	&BW	(1)& 2.87& 3.35&$0.018^{+0.001}_{-0.001}$	&$0.097^{+0.001}_{-0.001}$ &$27.93^{+1.18}_{-0.82}$	&33.75	&1,5 \\
4 &PSR J0952$-$0607	&BW	(1)& 1.41& 4.56&$0.029^{+0.001}_{-0.001}$	&$2.043^{+0.077}_{-0.076}$ &30.47	&34.81	&1,7,8,9 \\
5 &PSR J1124$-$3653	&BW	(1)& 2.41& 1.41&$0.041^{+0.001}_{-0.003}$	&$0.286^{+0.008}_{-0.010}$ &$30.56^{+0.45}_{-0.68}$	&33.60	&1 \\
6&PSR J1301$+$0833	&BW	(1)& 1.84 &-&$0.045^{+0.001}_{-0.002}$	&$0.697^{+0.046}_{-0.058}$ &-	&34.82	&1,10  \\
7&PSR J1311$-$3430	&BW	(1)& 2.56& 20.92&$0.016^{+0.001}_{-0.001}$\textsuperscript{(b)}	&$\sim$20 &$31.63^{+0.37}_{-0.54}$	&34.69	&11,12 \\
8&PSR J1544$+$4937	&BW	(1)& 2.16& 2.93&$0.025^{+0.006}_{-0.006}$ &$0.161^{+0.035}_{-0.035}$ &-	&34.06	&15,16 \\
9&PSR J1555$-$2908  &BW (1)& 1.79& 44.5&$0.060^{+0.005}_{-0.003}$\textsuperscript{(d)} &$9.92^{+0.002}_{-0.002}$\textsuperscript{(d)} &- &35.49 &3,43\\
10&PSR J1653.6$-$0158	&BW	(1)& 1.97& 2.40&$0.013^{+0.001}_{-0.001}$\textsuperscript{(e)}	&$0.333^{+0.039}_{-0.034}$\textsuperscript{(e)} &$31.07^{+0.04}_{-0.04}$	&33.64	&17,18,51 \\
11&PSR J1810$+$1744	&BW	(1)& 1.66& 4.60&$0.071^{+0.027}_{-0.027}$\textsuperscript{(k)}	&2.389 &$30.68^{+0.44}_{-0.52}$	&34.60	&19,2 \\
12&PSR J1959$+$2048	&BW	(1)& 1.61& 10.63&$0.036^{+0.001}_{-0.001}$	&$3.01^{+0.036}_{-0.036}$ &$31.04^{+0.32}_{-0.48}$	&35.00	&1,20,21   \\
13&PSR J2017$-$1614	&BW	(1)& 2.31& 2.45&  -\textsuperscript{(c)}	&- &-	&33.85	& 22  \\
14&PSR J2051$-$0827	&BW	(1)& 4.51& 12.13& $\sim$0.04	&- &$28.62^{+0.55}_{-0.70}$	&33.72	&23,24 \\
15&PSR J2052$+$1219	&BW	(1)& 1.99& 6.70&$0.042^{+0.001}_{-0.001}$	&$0.944^{+0.024}_{-0.024}$ &-	&34.53	&1,25 \\
16&PSR J2241$-$5236	&BW	(1)& 2.19& 6.65&$0.016^{+0.001}_{-0.001}$	&$0.282^{+0.005}_{-0.006}$ &$29.88^{+0.42}_{-0.57}$	&34.40	&1,26 \\
17&PSR J2256$-$1024	&BW	(1)& 2.29& 12.10& $>$0.030	&0.277 &$30.08^{+0.37}_{-0.53}$	&34.60	& 2 \\
18&PSR J1023$+$0038	&RB	(1)& 1.96& 11.81&$0.205^{+0.008}_{-0.008}$\textsuperscript{(j,f)}	&0.884 &$31.90^{+0.31}_{-0.75}$\textsuperscript{(i)}	&34.40	& 27,29,2 \\
19 &PSR J1048$+$2339	&RB	(1)& 4.67& 30.37&0.24-0.35	&$0.759^{+0.210}_{-0.175}$ &$31.52^{+0.18}_{-0.18}$\textsuperscript{(h)}	&34.08	& 30,31,32 \\
20 &PSR J1227$-$4853	&RB	(1)& 1.69& 11.11&$0.17^{+0.01}_{-0.01}$\textsuperscript{(j,f)}	&$\sim$1.21\textsuperscript{(f)} &$32.12^{+0.31}_{-0.47}$\textsuperscript{(i)}	&34.96	&27,33,34 \\
21 &PSR J1306$-$40	&RB	(1)& 2.20& - &$0.59^{+0.01}_{-0.01}$\textsuperscript{(g)}	&$\sim$5.90\textsuperscript{(g)} &31.94	& -	&35,36,47 \\
22 &PSR J1431$-$4715 	&RB	(2)& 2.01& 	14.11&0.13 - 0.19	&- &-	&34.83	&37,38 \\
23 &PSR J1622$-$0315	&RB	(2)& 3.85& 	11.6& $0.122^{+0.007}_{-0.006}$	&- &$30.61^{+0.07}_{-0.07}$\textsuperscript{(h)}	&33.89	& 39,40,41\\
24 &PSR J1628$-$3205	&RB	(2)& 3.21& 	15.06&0.17 - 0.24	&- &$30.96^{+0.33}_{-0.50}$	&34.26	& 10\\
25 &PSR J1723$-$2837	&RB	(2)& 1.86& 	7.56&$0.36^{+0.08}_{-0.06}$	&- &$31.92^{+0.31}_{-0.48}$	&34.67	&42,44,38\\
26 &PSR J2039$-$5617	&RB	(2)& 2.65& 	14.16&0.162 - 0.18	&$0.150^{+0.027}_{-0.018}$ &$31.18^{+0.03}_{-0.03}$\textsuperscript{(h)}	&34.40	&38,28,13,49\\
27 &PSR J2129$-$0429	&RB	(2)& 7.62& 	335.6&$0.44^{+0.04}_{-0.04}$	&$0.091^{+0.018}_{-0.018}$ &$31.79^{+0.34}_{-0.51}$	&34.48	&45,46 \\
28 &PSR J2215$+$5135	&RB	(1)& 2.61&  33.37&$0.345^{+0.008}_{-0.007}$\textsuperscript{(k)}	&0.794 &$31.92^{+0.41}_{-0.61}$	&34.87	& 19,2\\
29 &PSR J2339$-$0533 &RB	(1)& 2.88&  6.68&$0.30^{+0.02}_{-0.02}$\textsuperscript{(l)}	&$0.316^{+0.073}_{-0.073}$ &$31.44^{+0.33}_{-0.49}$	&34.04	&48,50 \\
\hline
    \end{tabular}
     \small\scalebox{1}{}{ 
    \begin{flushleft}
    \textbf{References.} (1) \cite{Draghis2019}, (2) \cite{Breton2013}, (3) \cite{Kennedy2022}, (4) \cite{Deneva2021}, (5) \cite{Stovall2014}, (6) \cite{Spiewak2016}, (7) (0.3-10 keV) \cite{Ho2019}, (8), \cite{Bassa2017}, (9) \cite{Nieder2019}, (10) \cite{Li2014}, (11) \cite{Romani2012}, (12) \cite{Ray2013}, (13) \cite{Clark2021}, (14) (2-10keV) \cite{Lee2018}, (15) \cite{Tang2014}, (16) \cite{Bhatta2013}, (17) \cite{Kong2014}, (18) \cite{Nieder2020}, (19) \cite{Schroeder2014}, (20) \cite{Guillemot2012}, (21) \cite{Huang2012}, (22) \cite{Sanpa}, (23) \cite{Stappers2001}, (24) \cite{Espinoza2013}, (25) \cite{Zharikov2019}, (26) \cite{Keith2011}, (27) \cite{Stringer2021}, (28) (0.5-8 keV) \cite{Bogdanov2021}, (29) \cite{Bogdanov2011}, (30) \cite{Deneva2016}, (31) (0.3-10 keV) \cite{Zanon2021}, (32) \cite{Yap2019}, (33) \cite{Roy2015}, (34) \cite{Bogdanov2014}, (35) \cite{Swihart2019}, (36) (0.5-10 keV) \cite{Linares2018}, (37) \cite{Bates2015}, (38) \cite{Strader19}, (39) this work, (40) \cite{Sanpa}, (41) (0.2-10 keV) \cite{Gentile2018PhD}, (42) \cite{Crawford2013}, (43) \cite{Ray2022}, (44) \cite{Staden2016}, (45) \cite{Bellm2016}, (46) \cite{Hui2015}, (47) \cite{Keane2018}, (48) \cite{Kandel2020Apj}, (49) \cite{Romani2015}, (50) \cite{Romani2011}, (51) \cite{Long_2022}. \\ \textbf{Notes.}\\ 
    -$\dot{P}, \tau, L_{\rm x}, \dot{E}$ are taken from reference (14) and the references therein, unless otherwise stated. \\
    -Irradiation power is calculated using $L_{\rm irr}= 4\pi a^2\sigma({T_2}^{4} - {T_1}^{4})$ or $L_{\rm irr}=\epsilon\dot{E}$, depending on the available information.\\
    Items: $^{(a)}$ Number of optical peak(s). $^{(b)}$ Parameters from the `Eq Hot Spots' model are used. $^{(c)}$ No light curve derived parameters. $^{(d)}$ 1$\sigma$ value is used. $^{(e)}$ Parameters from the `Veiled model' are used. $^{(f)}$ Parameters from the `HR2' model are used; $a$ = 0.97$\;\! {\rm R}_\odot$ obtained from reference (33). $^{(g)}$ Temperatures from the `No spot' model are used; $a$ is calculated using $T = 2\pi\sqrt{a^3/G/(M_1+M_2)}$. $^{(h)}$ $\Delta(\log L_{\rm x})=(\Delta L/L)/\ln{10}$. $^{(i)}$ Rotation-powered state. $^{(j)}$ Companion mass is calculated using the reported pulsar mass and binary mass ratio. $^{(k)}$ Parameters from the `NextGen' model are used. $^{(l)}$ Parameters from the `HS' model are used.
    \end{flushleft} 
    }
    
    \caption{A compilation of spider MSPs with published optical counterparts.}
    \label{tab:profile}
\end{table*}



\section{Pulsar irradiation and other system properties}
\label{sec:MWL}
The interplay between the energy loss of a neutron star that gives rise to radiation and the mass transfer from the companion (which, in turn, increases the rotational energy of the pulsar) determines the evolution state of a NS low-mass X-ray binary. 
Many dedicated observations have been conducted in the last decade to study the multi-wavelength properties of BWs and RBs.
However,
the transient nature of these systems increases the uncertainty in constraining binary evolution models.  
An empirical method is often used to understand the behavior of different astronomical systems.
We study the relationship between 
the
irradiation luminosity ($L_{\rm irr}$), the X-ray luminosity ($L_{\rm x}$), and the spin-down luminosity ($\dot{E}$)
 in BWs, RBs and tMSPs using an empirical method.
The relationship between the X-ray properties and spin-down luminosities of BWs 
  and RBs has been previously investigated \citep[see, e.g.][]{Arumu2015, Lee2018}.
  In particular, using a simple power-law model, \cite{Lee2018} showed that the X-ray emission of RBs is brighter and harder compared to that of BWs. 
Among the spiders, tMSPs are brighter X-ray sources ($\sim\!10^{32}$\;\!erg\;s$^{-1}$ during rotation-powered state; or see \citealt{Li2020}).

In general, for optical modeling of BWs and RBs, the pulsar irradiation is 
  computed as 
$L_{\rm irr}\propto ({T_2}^{4} - {T_1}^{4})$, where $T_1$ and $T_2$ are often the base or night-side temperature and the day-side temperature of the companion.
The companion is usually a low-mass star, and, depending on its spectral type, 
  the increased temperature due to the pulsar's irradiation on its inner face ($\Delta T = T_2 - T_1$) can be more than a few thousand K for some BW systems.

We compiled a list of BWs and RBs in the Galactic field (c.f. Table 
  \ref{tab:profile}), for which optical light curves have been previously reported.
  We conducted a literature review and compiled the reported irradiation luminosities. If either the spin-down energy conversion efficiency or the characteristic irradiation temperature is reported, we
  computed the irradiation luminosities as described in the footnotes. 
We dedicated a column to record the number of peaks observed in the 
  optical light curve of each source. A tidally distorted companion without strong irradiation from the pulsar shows a two-peaked ellipsoidal light curve variation. This allows us to quickly identify the systems that are dominated by pulsar irradiation, which would otherwise show a 1-peak modulated light curve.
All BWs in our sample have 1-peak modulation.
We note that the detection of BW companions is more challenging since they are fainter, 
  and we may not be able to detect them if there are less irradiated.
The RBs in our sample have 1 or 2 peaks, or a combination of these two effects.
The X-ray luminosity and spin-down luminosity are also included in the table;
these values, 
where available, are based on the study of \cite{Lee2018}.
We also added the companion mass, spin and spin-down period to the table for completeness.
The tMSP parameters in the table correspond to the radio pulsar state.

\begin{figure*}
    \centering
    \includegraphics[scale=0.5]{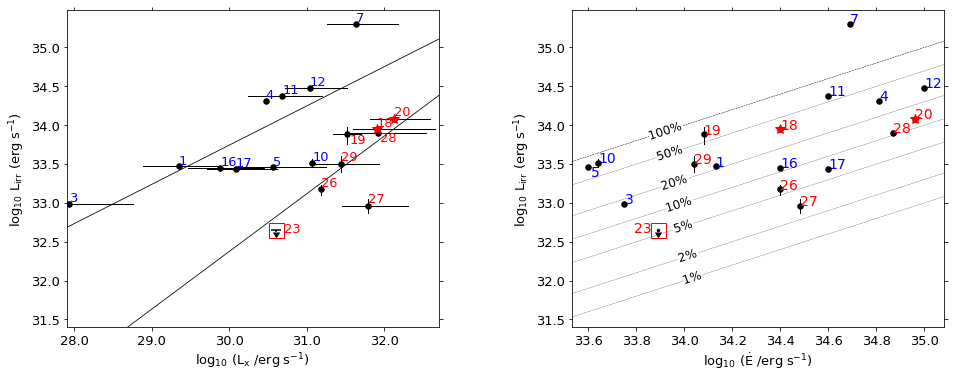}
    \caption{(Left) A comparison between the irradiation luminosity and the X-ray luminosity. (Right) A comparison between the irradiation luminosity and the spin-down luminosity. The ID of BWs are labelled in blue while RBs are labelled in red. tMSPs (ID: 18, 20) are marked as red stars. The upper limit of irradiation luminosity of PSR J1622-0315 (marked by a red square box) is calculated using the effective temperature. The RBs and BWs occupy two distinct regions in the $L_{\rm irr}$-$L_{\rm x}$ plot. The dotted lines in $L_{\rm irr}$-$\dot{E}$ plot correspond to different $L_{\rm irr}/\dot{E}$ ratios. PSR J1311$-$3430 (ID: 7) is the only source in our study that has $L_{\rm irr} > \dot{E}$.}
    \label{fig:ox}
\end{figure*}

As mentioned in the Introduction, spiders systems are believed to have evolved from LMXB systems. Therefore, it is useful to compare the X-ray properties of these systems to the irradiation luminosity experienced by the companions. 
Such a comparison allows us to evaluate the total energy available, and the evolution of the luminosity in these systems that suffer
constant ablation \citep{Levinson1991ApJ}.
The left panel of Figure \ref{fig:ox} shows the relationship between irradiation luminosity and X-ray luminosity. 
The irradiation luminosity is of about
2 orders of magnitude higher than the X-ray luminosity, 
and therefore the X-ray emission, regardless of its origin, is unlikely to be the main source of irradiation in these systems.
A RB system with higher X-ray luminosity could be experiencing weak accretion activity at some stages; on the other hand, BW systems are dominated by the pulsar irradiation, hence they are not accreting systems.
We performed a linear regression analysis for the two variables using the \texttt{scipy} package \texttt{linregress}. The Pearson correlation coefficients are 0.56 and 0.75, for RBs and BWs respectively. 
The two variables are related by $\log_{10} L_{\rm irr} = 0.74~\log_{10} L_{\rm x} + 10.11$ for RBs, and $\log_{10} L_{\rm irr} = 0.50~\log_{10} L_{\rm x} + 18.60$ for BWs.
 In our samples, all four highly irradiated BWs (ID: 4, 7, 11, 12) have high X-ray and spin-down luminosity compared to other BWs.
 The irradiation luminosity gap observed in the $L_{\rm irr}$-$L_{\rm x}$ plot for BWs (i.e. $0.3 \times 10^{34}$ erg\;\!s$^{-1}$$\lesssim L_{\rm irr, BW} \lesssim 2\times 10^{34}$ erg\;\!s$^{-1}$) corresponds to characteristic irradiation temperatures of 5400 K $\lesssim T_{\rm irr} \lesssim$ 8700 K, assuming an orbital separation of one solar radius. 
 By applying the empirical relation we obtained for BW systems and substituting for the rough upper and lower bounds of the observed gap, we found that the resulting X-ray luminosity is $L_{\rm x} \in (6\times 10^{29},2\times 10^{31}) $ erg\;\!s$^{-1}$. In the plot, we show that the X-ray counterpart of some of the BWs falls within this range, suggesting that the irradiation luminosity gap is likely not an observational bias.
We note that the derived irradiation temperature is dependent on the model used, and is 
  subject to the variable nature of spiders (e.g. \textcolor{black}{the pulsar heating in PSR J1048$+$2339 became significant in less than two weeks; \citealt{Yap2019}}), however this should not affect the general trend reported.
The tMSPs are the brightest among the RBs in both irradiation and X-ray luminosity. 

The right panel of Figure \ref{fig:ox} shows the relationship between irradiation luminosity and spin-down 
  luminosity.
The order of magnitudes between the two luminosities are similar ($\sim\! 
  10^{33}-10^{35}$\;\!erg\;s$^{-1}$).
  For PSR J1622-0315, while the system is dominated by ellipsoidal distortion, we cannot rule out the presence of pulsar irradiation. Therefore, we present the irradiation luminosity of PSR J1622-0315 as an upper limit in the same figure, where the effective temperature is used as the characteristic irradiation temperature. The error bar corresponds to 1$\sigma$ uncertainty from the fitting results.
Given that a system requires more energy to raise the temperature of a hotter star compared to a cooler star, and the effective temperature in PSR J1622-0315 is relatively higher among other spider companions, an irradiation temperature of $T_{\rm irr}\approx$ $T_{\rm eff}$=$6400~{\rm K}$ corresponds to a day-side temperature ${T_2}$=$7300$ K (if $T_1$=$6000$ K). A considerable amount of energy is required to achieve this temperature difference, hence the effective temperature is used as an approximation to estimate the upper limit of pulsar irradiation in PSR J1622-0315.

\begin{figure}
    \centering
    \includegraphics[scale=0.5]{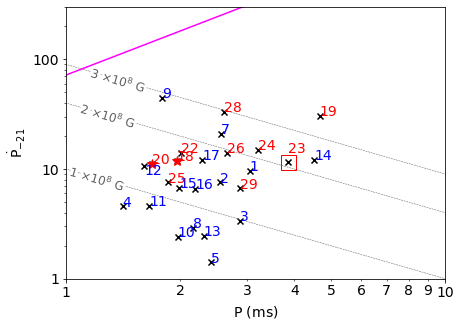}
    \caption{A complimentary $P$-$\dot{P}$ diagram of the pulsars in spider systems shown in logarithmic scale where $\dot{P}_{-21} = \dot{P}/10^{-21}$. The characteristic magnetic field strengths are indicated by dashed gray lines. The ID of BWs are labelled in blue while RBs are labelled in red. tMSPs are marked as red stars. The magenta line shows the MSP spin-up line according to \cite{Liu2022ApJ}.
    }
    \label{fig:opdot}
\end{figure}

In analogy to the 
  conversion efficiency from spin-down power to X-ray luminosity \citep[Fig. 10 in][]{Arumu2015} and $\gamma$-ray luminosity \citep[Fig. 11 in][]{Strader19},
 we added analytic lines that correspond to $L_{\rm irr}/\dot{E}$ for comparison.
 A conversion efficiency would suggest that 
 the heating mechanism involves some form of reprocessing of the spin-down energy. A theoretical study of this would be useful, but beyond the scope of this paper. The $L_{\rm irr}$-$\dot{E}$ relationship for BWs and RBs are similar in that they scatter across the parameter space despite the companion stars having masses well separated into two groups. 
 However, one of the BWs, PSR J1311-3430 has an 
irradiation luminosity that exceeds its spin-down luminosity ($L_{\rm irr} > \dot{E}$; conversion efficiency greater than 100$\%$), in contrast to other systems that have $L_{\rm irr} < \dot{E}$.
 It is likely that a less irradiated BW system would have a low conversion efficiency of only a few percent, and would be faint and challenging to observe.
To complement the $L_{\rm irr}$-$\dot{E}$ plot, in Figure \ref{fig:opdot} we show the $P$-$\dot{P}$ diagram of the pulsars in known spider systems.
The spin periods ($P$) of the pulsar in RBs and BWs are similar, but the pulsars in a RB system occupy the space with higher $\dot{P}$ values compared to the pulsars in a BW system. 
We note that the
tMSPs (marked red stars) have very similar $P$ and $\dot{P}$ values. 
We added the MSP spin-up line, reproduced from a recent study reported by \cite{Liu2022ApJ}. This information is useful for comparing the spin profile of irradiated transitional systems, which is outside the scope of this paper.



\section{Summary}
\label{sec:summary}
PSR J1622$-$0315 was discovered in the pulsar search of \textit{Fermi} unassociated sources 
  using the GBT
  and the optical counterpart was identified \citep{Sanpa}. The light curve showed a clear orbital variation and fluctuations at the same orbital phase.
\cite{Strader19} published the radial velocity and spectra of PSR J1622$-$0315, and the 
  companion mass was constrained to 0.10$-$0.14~$\textup{M}_\odot$.

In this work, we modeled the light curve of PSR J1622$-$0315 using the {\tt PHOEBE} code.  We 
  performed MCMC sampling on six system parameters and
found that the pulsar irradiation on the companion is not significant in this system 
  (c.f. Fig. \ref{fig:p1}).
The result obtained constrained the pulsar mass to $1.84^{+0.19}_{-0.19}~\textup{M}_\odot$ 
  and the companion mass to $0.122^{+0.007}_{-0.006}~\textup{M}_\odot$.
From the MCMC result, we find that PSR J1622$-$0315 is the lightest known redback system 
  with a relatively hot companion. 
We note that PSR J1622$-$0315 has an orbital period of 0.16~d, which is slightly smaller 
  than other RBs ($\gtrsim$ 0.2~d). 

Additionally,
we studied the energetic relationship between irradiation luminosity ($L_{\rm irr}$), X-ray luminosity ($L_{\rm x}$) and the pulsar's spin-down luminosity ($\dot{E}$) of spider systems. 
Both BWs and RBs show a correlation in their $L_{\rm irr}$ and $L_{\rm x}$ values, with BWs having a higher  
Pearson correlation of 0.75. 
We find that PSR J1622$-$0315 lies well within the RB group with insignificant pulsar irradiation. 
PSR J1622$-$0315 also
has the smallest X-ray and spin-down luminosity among the RBs in the sample.
Observations of borderline-mass (companion) spiders like PSR J1622$-$0315 and PSR J1810$+$1744 \citep{Schroeder2014} are important to constrain the binary evolution model.
Recent findings suggest that both NS low-mass X-ray binaries and spider systems tend to have an ionized and clumpy environment \citep{Knight2023MNRAS}, implying that
pulsar irradiation and mass loss in the form of winds are common in different evolution stages.
Assuming the presence of pulsar irradiation, 
it is useful to compare the companion star behavior to the neutron star properties as they co-evolve with each other.
For instance, in this study, both tMSPs are the RB systems with the highest pulsar irradiation luminosity and X-ray luminosity compared to other spider systems, and this suggests that the transitional systems are in general more energetic. 
Provided the system is bright enough to observed, both light curve modeling of the irradiation properties and spectral observation can be used together to study the evolution of the companion under pulsar irradiation.
Finally, the estimation of irradiation luminosity is model-dependent and is subject to the 
  variable nature of these systems, which may not be fully reflected in this study.


\begin{acknowledgements}
       We thank the referee for his/her many useful scientific suggestions and comments. This work has made use of data collected at Lulin Observatory, partly supported by the National Science and Technology Council of the Republic of China (Taiwan) through grant 105-2112-M-008-024-MY3. Y.X.J.Y and A.K.H.K. are supported by the National Science and Technology Council of the Republic of China (Taiwan) through grants 110-2628-M-007-005 and 111-2112-M-007-020. 
       KLL is supported by the Ministry of Science and Technology of the Republic of China (Taiwan) through grant 110-2636-M-006-013, and he is a Yushan (Young) Scholar of the Ministry of Education of the Republic of China (Taiwan).
This work used high-performance computing facilities operated by the
Center for Informatics and Computation in Astronomy (CICA) at National
Tsing Hua University. This equipment was funded by the Ministry of
Education of Taiwan, the Ministry of Science and Technology of Taiwan,
and National Tsing Hua University.
We thank Kinwah Wu and Kaye Li for their advice to improve the readability of the manuscript.
    
\end{acknowledgements}

\bibliography{ref}
\bibliographystyle{aasjournal}



\end{document}